# Correspondence

## Propagation Measurements and Modeling for Low Altitude UAVs From 1 to 24 GHz

Cesar Briso, Cesar Calvo, Zhuangzhuang Cui, Lei Zhang, and Youyun Xu

*Abstract*—In most countries, small (<2 kg) and medium (<25 kg) size unmanned aerial vehicles (UAVs) must fly at low altitude, below 120 m, and with permanent radio communications with ground for control and telemetry. These communications links can be provided using 4G/5G networks or dedicated links, but in either this case the communications can be significantly degraded by frequent Non Line of Sight (NLoS) propagation. In this case, reflection and diffraction from ground objects are critical to maintain links, and hence accurate propagation models for this must be considered. In this letter we present a model for path loss when the UAV is flying in NLOS conditions. The study is based on measurements made at frequencies of 1, 4, 12, and 24 GHz with a UAV flying in a suburban environment. Measurements have been used to model NLOS propagation below 4 GHz, where the dominant mechanism is diffraction, and above 4 GHz where multipath is the dominant propagation mechanism. The model can be of use in predicting excess losses when UAVs fly in suburban NLOS conditions.

## I. INTRODUCTION

The use of UAVs for commercial applications is increasing notably worldwide, and strong growth is expected in the coming years. Most commonly used UAVs are those having a maximum flying weight below 25 kg. These vehicles are allowed to fly below commercial aircraft, with a maximum altitude of 120 m (400 feet), in what is known as uncontrolled or non-segregated airspace. The flying range usually goes to a few hundred meters with visual line of sight (VLOS), to several km in operations called beyond visual line of sight (BVLOS).

With these limitations, small UAVs typically fly close to terrain with altitudes from 15 to 50 m, enough to overfly ground objects and to minimize the effect of the wind. In these flying conditions, the UAVs often do not have Line of Sight (LoS) communications with their control station, so that multipath and diffraction significantly affect propagation of UAV radio signals.

A commercial UAV requires two or three radio links: control, telemetry, and payload; sometimes control and telemetry are combined [1]. Control and telemetry are two functions necessary for the correct (and safe) operation of the UAV. Telemetry links often include a standard quality video link called "First Person View (FPV)" to provide information on the environment to the UAV pilot and to enable the safe operation of the UAV even beyond the pilot's line of sight. The payload link is typically a high capacity link for transmitting high quality video or other sensing signals for commercial or other applications.

All the radio links can be provided using 4G/5G technology based on terrestrial ground stations with antennas located atop towers (∼10–30 m). The use of these networks must be reliable under a variety of propagation conditions, including LoS and non-LoS (NLoS), with multipath components (MPCs), shadowing, and diffraction from ground objects. Propagation conditions are strongly influenced by the altitude of the UAV and the height of the ground station antenna. Multipath can be a strong function of altitude and environment as described in [2]. Similarly, interference changes dramatically with altitude [3]. Path loss and multipath models for medium altitude are given in [4], and for one low altitude setting in [5]. However, propagation when the ground station is on a telecommunications tower and vehicles fly close to building roof tops will be highly influenced by NLoS conditions, so it is necessary to estimate the path losses [6] in realistic conditions for UAVs, and to analyze the influence of multipath and diffraction together for a range of frequencies.

The diffraction measurements for radio channels have been thoroughly investigated in low-frequency bands in [7]–[9]. In [7], diffraction loss measurements of long distances about 5 km at a top of a mountain at 0.1–10 GHz are provided, which leads to a very small diffraction angle within 3 degrees. In [8], the comparisons between the diffraction measurements and the uniform geometrical theory of diffraction (UTD) were conducted at 2.4 GHz. In both cases the agreement with the diffraction ITU model is good. In [9], the diffraction measurement results for a building corner at 10 GHz were used to compare with a theoretical model of knife-edge diffraction (KED). The results show a good performance to predict the diffraction loss. For diffraction at frequencies below 4 Ghz, there is no new diffraction model proposed since the KED and UTD model can be used. Regarding the frequency band requirements of 5G, various measurements for the centimeter wave (cmWave) and millimeter wave (mmWave) bands were carried out in [10]–[13]. The outdoor and indoor 5G diffraction measurements were conducted at 10 GHz, 20 GHz and 26 GHz in [10] where the authors found that the diffraction loss for an outdoor building corner with rounded edges can be better predicted by a simple linear model with a fixed anchor point rather than the KED model. For potential bands in 5G, the measurements were conducted at 28 GHz in an over-rooftop scenario [11] and building corner scenarios [12]. Similar to [10], a linear model as a function of the diffracted angle was presented in [11]. However, the authors in [12] found that the diffraction loss presents a logarithmic change when the angle is small (0.1-5°) whereas it presents a linear change with larger angles (5-40°). When the angle is smaller than 0.1°, the loss is a constant (6.5 dB). Thus, a segmented diffraction model was proposed for different diffraction angles. Generally, the diffraction loss at low frequencies can be predicted by classical models such as KED and UTD. At higher frequency, the linear and logarithmic models are commonly used.

In this letter we describe results of measurements and subsequent modeling of diffraction and multipath under NLoS in realistic

Manuscript received May 29, 2019; revised November 8, 2019; accepted January 14, 2020. Date of publication January 22, 2020; date of current version March 12, 2020. This work was supported in part by the Chinese Strategic International Cooperative Project of National key R&D Plan under Grant 2016YFE0200200, and in part by the National Nature Science Foundation of China (NSFC) Project under Grant 61901104. The review of this article was coordinated by Dr. J. Paris. *(Corresponding author: Youyun Xu.)*

C. Briso, C. Calvo, and Z. Cui are with the Technical University of Madrid, 28040 Madrid, Spain (e-mail: cesar.briso@upm.es; cesar.calvo.ramirez@alumnos.upm.es; 16120053@bjtu.edu.cn).

L. Zhang is with the Donghua University, Shanghai 200000–202100, China (e-mail: lei.zhang@dhu.edu.cn).

Y. Xu is with the Nanjing University of Post and Telecomunications, Jiantsu, China (e-mail: yyxu@njupt.edu.cn).

Digital Object Identifier 10.1109/TVT.2020.2968136

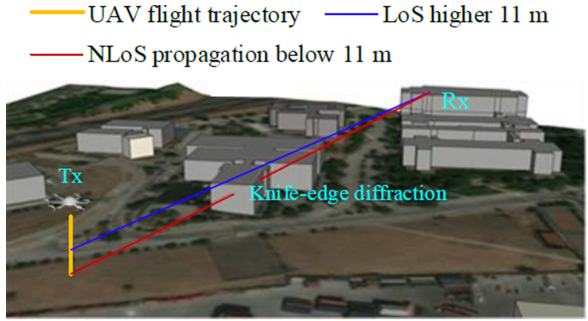

Fig. 1. Detailed view of the test environment where the NLoS propagation is below 11 m height.

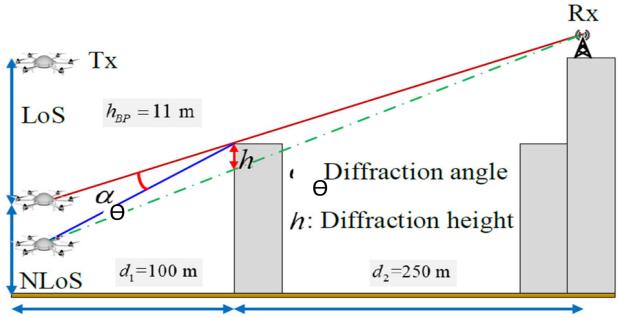

Fig. 2. Detailed view of the measurement area and cross section of the diffraction building where $\theta$ and $h$ are the diffracted angle and height, respectively.

UAV flight conditions. The modeling includes validation at different frequencies <4 GHz (1000, and 4000 MHz), where the diffraction is dominant and two frequencies above >4 GHz (12 and 24 GHz) where the dominant propagation mechanism is multipath. The ITU diffraction model is used, combined with the modeling of the multipath power based on the measurements of received power for LoS and NLoS conditions at high frequencies. Comparisons with other models are also provided in Section V.

## II. TEST ENVIRONMENT AND PROPAGATION MEASUREMENTS

The modeling of propagation in NLoS conditions requires a controlled environment where propagation conditions of LoS and NLoS can be clearly defined. We have carefully chosen a test environment where the diffraction point is clearly defined and the use of 4G/5G base stations for UAV communications has been emulated by installing the test receiver on the top of a building.

### A. Test Environment

The test area is located in a suburban environment at the Technical University of Madrid, Spain, with several buildings with an average height of 15 m. The test receiver (Rx), is installed on the top of one of the highest buildings at a height of 25 m. The UAV ascended from an open field area at distance 350 m from the Rx ground station (GS). There is a 15 m height building blocking the transmitter and receiver that is used to induce LoS/NLoS conditions. This building is at 100 m from the UAV and 250 m from the GS. A detailed view of the test environment is shown in Fig. 1. A cross-section view appears in Fig. 2.

During takeoff for the first few meters of ascent the UAV is in NLoS conditions. When it reaches an altitude of 11 m it begins to have LoS with the GS. Once the UAV is well above the building it gains a clear LoS to the Rx GS (clearing more than 0.6 of the first Fresnel zone), thus, this configuration allows us to accurately measure the influence of diffraction and multipath on the air to ground link in NLoS conditions.

TABLE I
FLIGHT TEST DATA

| | TEST TRANSMITTER & RECEIVER | | |
|---|---|---|---|
| Freq. MHz | Transmitter power (EIRP,dBm)/ antenna gain | Test Receiver RX antenna Gain / PathLoss$_{LoS}$ | Measurement Dynamic Range (DR) |
| 1000 | 32.2 / 2.15 dBi | 2.15 dBi / 79 dB | 83.2 dB |
| 4000 | 31.3 / 2.15 dBi | 2.15 dBi / 91 dB | 70.3 dB |
| 12000 | 22.2 / 2.15 dBi | 8.5 dBi / 94.1 dB | 58.1 dB |
| 24000 | 13.5 / 2.15 dBi | 8.5 dBi / 100.1 dB | 43.4 dB |
| UAV | | | |
| Type | | | Hexacopter |
| Flight controller | | | DJI N3 with flight data logging |
| Flying weight with test transmitter | | | 4.7 Kg |
| Location of antennas | | | Lower part |
| Polarization | | | UAV Vertical |
| TEST ENVIRONMENT | | | |
| Height of diffraction object | | $h_D$ | 15 m |
| Distance of object to UAV | | $d_1$ | 100 m |
| Distance of object to GS | | $d_2$ | 250 m |
| UAV flying altitude | | $h_{UAV}$ | 0-30 m |
| Ground station height | | $h_{GS}$ | 25 m |
| Minimum altitude for LOS: | | $h_{BP}$ | 11 m |

### B. UAV Test Transmitter and Ground Station Receiver

The system used to perform the measurements is based on a hexacopter that carries a continuous wave transmitter. The UAV employs a state-of-the-art flight controller that provides a real-time recording of flight data, providing accurate 3D GNSS positioning and information on accelerations, altitude, time, and other flight parameters.

The test transmitters were custom made and were configured to operate at 1000, 4000 MHz, and at 12 and 24 GHz. The output power is given in Table I. At all frequencies omnidirectional $\lambda/4$ antennas were used in the UAV: a broadband ($\lambda/4$ equivalent) antenna model MGRM-WHF for the 1–6 GHz band, and low gain with wide elevation angle $\lambda/4$ monopoles for 12 and 24 GHz.

As noted, the GS was located on the top of a 25 m tall building at 350 m from the UAV. The GS receiver was a spectrum analyzer equipped with acquisition software that automatically records received power at a rate of 10 measurements/second. Receiver antennas were a 2.15 dBi $\lambda/4$ monopole with wide elevation beam for 1000–4000 MHz band, and an R&S HL050 8.5 dBi log periodic antenna for 12 and 24 GHz.

All antennas were calibrated in an anechoic chamber installed on the UAV, in order to guarantee a clean radiation pattern for elevation angles from 0° to 60° (Fig. 3).

The measurements were carefully designed to guarantee that the power received in NLoS conditions is well above the noise floor. For this purpose, we have computed the maximum received power (Pr$_{Max}$) in LoS condition for each frequency. Considering transmitter power (EIRP) and computing LoS path loss using the Friis equation for link distance $D$ ($D = d_1 + d_2$) between UAV and TX as defined in Fig. 2, we have,

$$Pr_{Max}(f) = EIRP(f) - PathLoss_{LoS}(D, f). \quad (1)$$

We then computed the Noise Floor (NF) considering that measurements were made with a 10 kHz bandwidth and using the preamplifier of the spectrum analyzer that has 4 dB maximum noise figure. With this configuration, the Noise Power (NP) was,

$$NP = -\frac{174 \text{ dBm}}{Hz} + 10\log_{10}(10,000 \text{Hz}) + 4 \text{ dB}$$

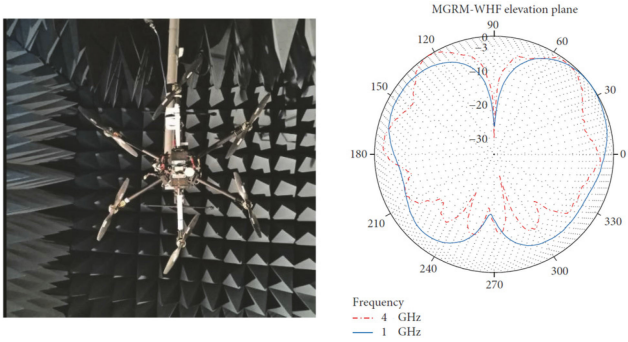

Fig. 3. Measurements of the elevation radiation pattern of antenna installed on UAV for 1–4 GHz.

$$= -130 \text{ dBm}. \quad (2)$$

Then, the Dynamic Range (DR) of measurement for each frequency is,

$$DR(f) = Pr_{Max}(f) - NP. \quad (3)$$

The DR describes the maximum path loss we can measure in NLoS conditions for each frequency. The lowest value is 43.4 dB at 24 GHz, i.e., more than 43 dB attenuation beyond LoS attenuation. In Table I we summarize all the parameters of the test.

## III. Propagation in NLoS

The objective of this section is to analyze and model propagation mechanisms below and above 4 GHz. Below 4 GHz, it is well known that a dominant mechanism in NLoS condition can be diffraction and the general diffraction model proposed by ITU [6] can be used to estimate diffraction loss. At higher frequencies, diffraction losses increase so that multipath propagation becomes more important and can become the dominant propagation mechanism, particularly at microwave (mW) and millimeter wave (mmW) frequencies.

As noted, measurements were made at several frequencies via vertical flights with the test UAV in the suburban environment illustrated in Fig. 2. In this environment there is one building that can be modeled as a knife edge to evaluate the diffraction losses; multipath is moderate in this locale, and has been previously characterized in [2].

### A. Diffraction Model

The most commonly used geometry for diffraction study is a single knife-edge diffracting object, here between UAV and GS. This knife-edge is an idealized abstraction of the roof top of a building. According to this diffraction model, the ITU proposed a recommendation [6] for excess loss using an analytical expression dependent on the geometry described in Fig. 2.

In (4) we define a "Breaking Point" height $h$ to distinguish LoS and NLoS propagation. This height is related to the altitude of the UAV and is used to compute the diffraction diffraction parameter $v$, defined as a function of the single knife-edge geometry and frequency (5). The expression $J(v)$ in (6) is the Fresnel-Kirchoff gain due to diffraction,

$$h = h_{BP} - h_{UAV}, \quad (4)$$

$$v = h\sqrt{\frac{2}{\lambda}\left(\frac{1}{d_1} + \frac{1}{d_2}\right)}, \quad (5)$$

$$J(v) = \left(\frac{\sqrt{[1 - C(v) - S(v)]^2 + [C(v) - S(v)]^2}}{2}\right), \quad (6)$$

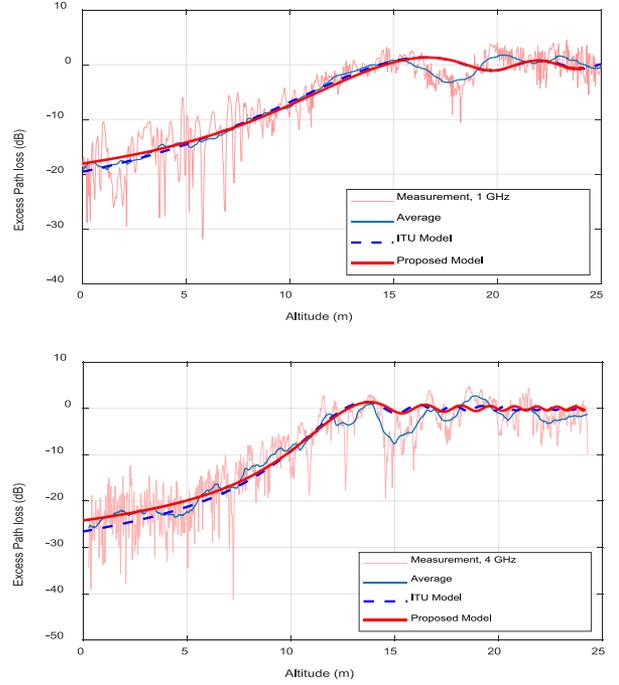

Fig. 4. Measurement and ITU model path loss vs. altitude comparison at 1 GHz and 4 GHz.

where $C(v)$ and $S(v)$ are the Fresnel cosine and sine integrals defined by,

$$C(v) = \int_0^v \cos\left(\frac{\pi s^2}{2}\right) ds, \quad (7)$$

$$S(v) = \int_0^v \sin\left(\frac{\pi s^2}{2}\right) ds. \quad (8)$$

These solutions can be obtained numerically, and results for the four frequencies as a function of altitude are presented in Figs. 3 and 4 as the ITU Model, defined as follows:

$$ITU\ Model = 20\log_{10}\left(\frac{1}{J(v)}\right). \quad (9)$$

From the diffraction model, we have optical LoS when $v = 0$ in (3). In this case diffraction loss is 6 dB, independent of frequency. In our case, this situation happens at an altitude of 11 m. Therefore, we have used this "Break Point" ($h_{BP}$) to distinguish between the LoS and NLoS conditions for modeling path loss.

### B. Measurement Results

As noted, during measurements the UAV made a vertical flight moving from NLoS condition, below $h_{BP} = 11$m, to LoS condition. During the flight a time reference was used to synchronize the received power measured at the GS, while the flight controller recorded the height of the UAV as a function of time. Measurements have been normalized and compared with the ITU diffraction model.

The results are shown in Figs. 3 and 4, where we have the actual measurement sample results along with a spatial average made with 100 samples to filter the small-scale fading and facilitate comparison between the ITU model and our path loss modeling.

Fig. 4 shows quite good agreement with the ITU diffraction model for all heights at the lower two frequencies of 1 and 4 GHz. The maximum

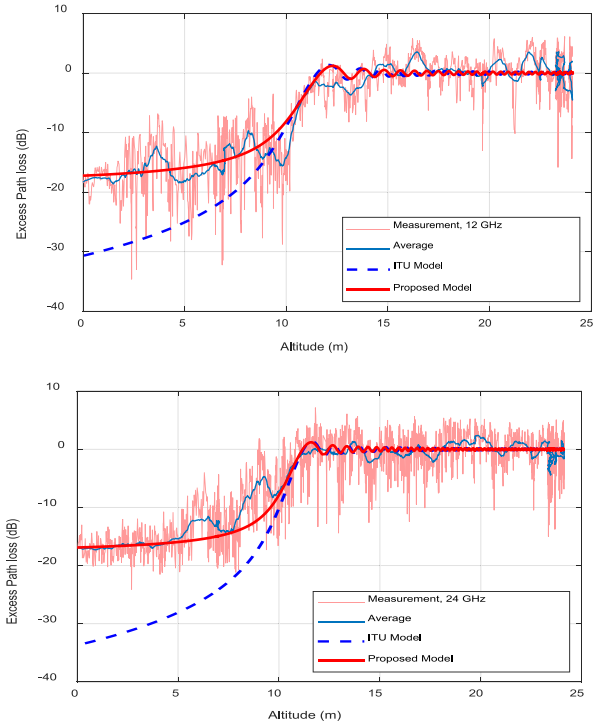

Fig. 5. Measurement and ITU model path loss vs. altitude comparison at 12 and 24 GHz.

RMS error between measurements and model is 1.9 dB for 1 and 4 GHz. Diffraction is thus the dominant propagation effect in this UAV to GS link.

Nevertheless, in Fig. 4 at 4 GHz we can see that from approximately 0 to 4 meter altitude, when diffraction losses are larger than 20 dB, the ITU mode l prediction is off by several dB, hence it is likely that some influence of multipath is contributing. This effect will be dominant at 12 and 24 GHz as is shown in Fig. 4.

## IV. MODELING EXCESS PATH LOSS IN NLoS CONDITION

As noted previously, the modeling of propagation in NLoS conditions must consider diffraction, multipath and scattering to accurately model path loss across frequency and for different UAV altitudes. We can see in Fig. 5 how the ITU model does not fit well for 12 and 24 GHz measurements, so that we have defined the following equation to model "Excess Path Loss" in LoS/NLoS conditions:

$$E_{PL}(h_{UAV}, f) = 20\log_{10}\left(1 + \frac{1}{\alpha(f)}\right) \\ - 20\log_{10}\left(\frac{1}{J(v)} + \frac{1}{\alpha(f)}\right). \quad (10)$$

In this equation, we use two coefficients to model $E_{PL}$: the "Diffraction losses" defined by the ITU model, $1/J(v)$, and a coefficient $1/\alpha(f)$ used to model the received power in NLoS conditions due to multipath and scattering from the environment. This coefficient can be obtained from measurements of Figs. 4 and 5, by fitting equation (10) for $E_{PL}$ at ground level. In this case we have $h_{UAV} = 0$, and $h = h_{BP}$ so that,

$$v_{BP} = h_{BP}\sqrt{\frac{2}{\lambda}\left(\frac{1}{d_1} + \frac{1}{d_2}\right)} \quad (11)$$

TABLE II
$E_{PL}$ AND ITU MODEL AT GROUND LEVEL; NLoS COEFFICIENT ($\alpha(f)$) FOR DIFFERENT FREQUENCIES

| Frequency | $h_{BP}$ | $E_{PL}$ (0,f) | ITU Model | $\alpha(f)$ |
|---|---|---|---|---|
| 1 GHz | | 17.9 dB | 19.5 dB | 60 |
| 4 GHz | 11 m | 24.2 dB | 26.5 dB | 35 |
| 12 GHz | | 17.2 dB | 30.7 dB | 7.9 |
| 24 GHz | | 16.9 dB | 33.7 dB | 6.9 |

and,

$$E_{PL}(0, f) = 20\log_{10}\left(1 + \frac{1}{\alpha(f)}\right) \\ - 20\log_{10}\left(\frac{1}{J(v_{BP})} + \frac{1}{\alpha(f)}\right), \quad (12)$$

where $E_{PL}(0, f)$ is obtained from measurements and $\frac{1}{J(v_{BP})}$ is a constant. Results for different frequencies of $E_{PL}$, ITU Model and $\alpha(f)$ are given in Table II. We can see that $1/\alpha(f)$ increases with frequency as a consequence that multipath and scattering becomes dominant over diffraction at higher frequencies. This happens because diffraction losses increased moderately with frequency while multipath/scattered power increase slowly [8].

On Figs. 4 and 5 we have represented the Excess Path Loss model fitted for the different frequencies with coefficients of Table II. At lower frequencies, Fig. 4, diffraction is dominant and the ITU model fits very well. Nevertheless, at higher frequencies we have a moderate level of multipath/scattering due to surrounding buildings, trees, cars and other objects, so that diffuse propagation becomes more relevant above 4 GHz and dominant at 12/24 GHz. On this case it is necessary to correct ITU model with $\alpha(f)$ parameter as shown on Fig. 4.

Therefore, we can summarize that to model an environment in NLoS conditions, we have to locate the position of the principal object blocking the LoS and measure the received power at fixed altitude. With these parameters we can compute the diffraction gain (DG) and the diffuse parameter $\alpha(f)$, and model the excess path loss of the environment for NLoS according to the height of the UAV.

## V. COMPARISONS WITH OTHER MODELS AND RANGE OF VALIDITY OF THE PROPOSED MODEL

To explain the improvements of our model we compare it with other diffraction models in this section. First, the ITU model based on KED has been validated at low frequencies (below 4 GHz), whereas in the high frequency, the model presents a poor performance in predicting the diffraction loss. The other models in the literature also support our conclusion [8]. Moreover, for high frequency bands, authors in [11] and [12] propose the linear model and logarithmic model, respectively, where the functional relationship refers to the diffraction angle. Since the measurements in [11] and [12] are carried out at 28 GHz (without considering the dependence of frequency), the comparison can be conducted for 24 GHz in this paper by adjusting the parameters of the model. Both models compute losses as a function of the diffracted angle ($\theta$ in Fig. 2). On our measurements, the diffracted angle is small ($0 \leq \theta \leq 6.24°$). Considering this, the linear model [11] is expressed as,

$$L(\theta) = \left(a\sqrt{(H - h_u)^2 + d_1^2} + b\right) \cdot \theta, \quad (13)$$

## VI. Conclusion

In this paper, we have presented results from a set of propagation measurements at four frequencies ranging from 1 to 24 GHz, and a model for excess path loss in a type of NLoS condition for a UAV communication link in a suburban environment. Comparing path loss results with diffraction theory, the single knife-edge diffraction model of the ITU has shown to be very accurate for frequencies below 4 GHz. At higher frequencies we found that the local diffuse propagation becomes dominant at 12 and 24 GHz. From this we developed a model that balances diffraction and multipath/scattering.

The proposed model it easy to fit knowing the position of the principal object blocking LoS and measuring diffuse power in NLoS at ground level. The model will be useful to improve the communication links with UAVs in such scenarios.

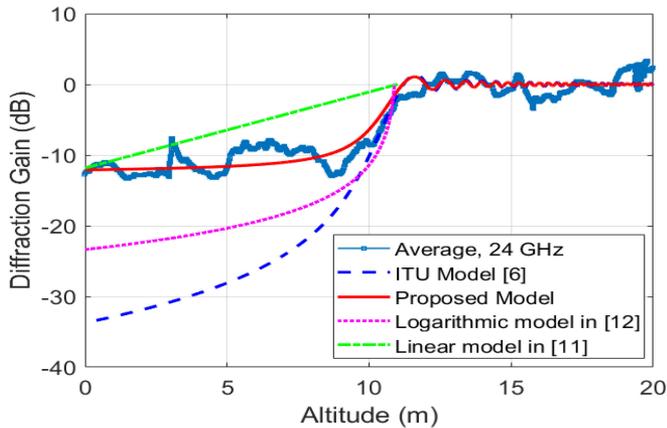

Fig. 6. Diffraction gain vs. altitude comparisons with other diffraction models using the measurement result and our proposed model at 24 GHz.

where $H$ and $h_u$ are the height of the diffracting building and UAV, respectively. Then $a$ and $b$ are linear coefficients. For our comparison, the values of $a$ and $b$ have been obtained by least-square method: $a = 0.011$ and $b = 0.785$.

The logarithmic model [12] for the range of 0.1–5° and can be expressed as:

$$L(\theta) = 5 \cdot \ln(\theta) + c \tag{14}$$

where $c$ is 18 dB at 28 GHz in [12] because on this measurements diffraction losses when $\theta < 0.1°$ are set to 6.5 dB. In our comparison, the diffraction losses when $\theta < 0.1°$ are equal to 0 dB so that $c$ is set to 11.5 dB to fit the curve. On Fig. 6 it can be shown that our proposed model has the best performance to predict the loss while the ITU model has the worst prediction among these models for NLOS conditions.

The linear model can predict the loss in the large diffracted angle (low altitude, such as 0 m) while the logarithmic model is more accurate in small angel when the altitude of UAV is close to LOS (9–11 m). Generally, our proposed model has a better performance to predict the loss mingling the MPCs and diffraction components.

The range of validity of the proposed model is defined by Rec. ITU-R P.526-14 [1] for single obstacle deterministic diffraction model. This recommendation defines a confidence interval for diffraction angle (Fig. 2) of $\Theta = 0$–12° for any distance at 10 MHz and above. This model is useful in any environment.

The deterministic multipath coefficient $\alpha(f)$ included to model multipath has been fitted on a range 1–24 GHz for a suburban environment. If the type of environment changes (Rural, Hilly, Urban, others) this coefficient has to be fitted again with additional measurements. These measurements can be narrow band received power measured on vertical flights. With these measurements we can fit path losses and obtain the parameter $\alpha(f)$ for a new environment.